\documentclass[aps,prl,amsmath,twocolumn]{revtex4}

\usepackage{amsmath}
\usepackage{bm}
\usepackage[dvips]{color,graphicx}
\usepackage{epsfig}
\usepackage{dcolumn}
\def\sw{\sin^2\theta_W}
\usepackage{latexsym}

\begin{document}

\title{New formulation of $\gamma Z$ box corrections to the weak charge
	of the proton}

\author{P. G. Blunden,$^1$ W. Melnitchouk$^2$ and A. W. Thomas$^3$}

\affiliation{
$^1$\mbox{Department of Physics and Astronomy, University of Manitoba}, 
	Winnipeg, MB, Canada R3T 2N2	\\
$^2$\mbox{Jefferson Lab, 12000 Jefferson Avenue, Newport News,
	Virginia 23606, USA}		\\
$^3$\mbox{CSSM, School of Chemistry and Physics, University of Adelaide},
	Adelaide SA 5005, Australia}

\begin{abstract}
We present a new formulation of one of the major radiative corrections
to the weak charge of the proton -- that arising from the axial-vector
hadron part of the $\gamma Z$ box diagram, $\Re{\rm e}\, \Box_{\gamma
Z}^{\rm A}$.
This formulation, based on dispersion relations, relates the $\gamma Z$
contributions to moments of the $F_3^{\gamma Z}$ interference structure
function.
It has a clear connection to the pioneering work of Marciano and Sirlin,
and enables a systematic approach to improved numerical precision.
Using currently available data, the total correction from all
intermediate states is $\Re{\rm e}\, \Box_{\gamma Z}^{\rm A}
= 0.0044(4)$ at zero energy, which shifts the theoretical estimate
of the proton weak charge from $0.0713(8)$ to $0.0705(8)$.
The energy dependence of this result, which is vital for interpreting
the Q$_{\rm weak}$ experiment, is also determined.
\end{abstract}

\maketitle

%%%%%%%%%%%%%%%%%%%%%%%%%%%%%%%%%%%%%%%%%%%%%%%%%%%%%%%%%%%%%%%%%%%%%%%%%

As modern parity-violating (PV) experiments press to ever improving
levels of precision, they remain a vital complement to direct tests
of the Standard Model at the high energy frontier.
The classic example of this, involving precise measurements of parity
violation in atoms, led to a remarkably accurate determination of
$\sin^2 \theta_W$.
A complementary PV electron-proton scattering measurement underway by
the Q$_{\rm weak}$ Collaboration~\cite{QWEAK} at Jefferson Lab has the
potential to increase the mass scale associated with new physics to
2~TeV or higher, provided that the critical radiative corrections are
under control.
In this Letter we present a new formulation of the important $\gamma Z$
radiative corrections which allows for their controlled, systematic
evaluation.

Including electroweak radiative corrections, the proton weak charge
is defined, at zero electron energy $E$ and zero momentum transfer,
as~\cite{Erler03}
\begin{eqnarray}
Q_W^p &=& (1 + \Delta\rho + \Delta_e)
	  (1 - 4 \sw(0) + \Delta_e')			\nonumber\\
      & & +\, \Box_{WW} + \Box_{ZZ} + \Box_{\gamma Z}(0)\, ,
\label{eq:Qwp}
\end{eqnarray}
where $\sw(0)$ is the weak mixing angle at zero momentum, and the
corrections $\Delta\rho$, $\Delta_e$ and $\Delta_e'$ are given in 
\cite{Erler03} and references therein.
% ###
The contributions $\Box_{WW}$ and $\Box_{ZZ}$ arise from the $WW$ and
$ZZ$ box and crossed-box diagrams, and can be computed perturbatively.
They are expected to be energy independent for electron scattering in
the GeV range.  By contrast, the $\gamma Z$ interference correction
$\Box_{\gamma Z}(E)$ depends on physics at both short and long distance
scales.

In the classic work of Marciano and Sirlin (MS) \cite{MS},
$\Box_{\gamma Z}(0)$ was evaluated in a quark model-inspired
loop calculation using either a ``perturbative'' (p) or a
``nonperturbative'' (np) {\it ansatz},
\begin{equation}
\Box_{\gamma Z}(0) = v_e(M_Z^2)\,{5\alpha\over 2\pi} \, B_{\rm p(np)},
\label{eq:boxgz}
\end{equation}
where
$v_e(M_Z^2) = (1-4\hat{s}^2)$, and
$\hat{s}^2 \equiv \sin^2{\theta_W(M_Z^2)} = 0.23116$
in the $\overline{\rm MS}$ scheme \cite{PDG}.

The perturbative {\it ansatz} \cite{MS}
\begin{equation}
B_{\rm p} = \ln{M_Z^2\over m^2} + \frac{3}{2}
\label{eq:MSp}
\end{equation}
is the free quark model result, with $m$ a hadronic mass scale,
and shows the leading-log behavior.
For the nonperturbative {\it ansatz}, $B_{\rm np} = K_m + L_m$
is the sum of a long-distance part, $L_m$, and a short-distance part,
$K_m$, with
\begin{equation}
K_m = \int_{m^2}^\infty {du\over u (1+u/M_Z^2)}
\left(1-{\alpha_s(u)\over\pi}\right).
\label{eq:MSnp}
\end{equation}
Here $m$ is a mass scale representing the onset of asymptotic 
behavior at large loop momenta, and the factor $(1-\alpha_s(u)/\pi)$
is the lowest-order correction induced by the strong interactions.
In Ref.~\cite{MS} $L_m$ is taken to be the elastic nucleon (Born)
contribution, which is evaluated to be 2.04 using the same dipole
form factors for both the electromagnetic and axial-vector coupling. 
MS \cite{MS} originally adopted the value $K_m = 9.6 \pm 1$, 
based on calculations with $m$ in the range 0.3--1.0~GeV.
A more recent calculation by Bardin {\it et al.}~\cite{Bardin} sets 
$0.5 \le m \le 0.6$~GeV, over which $K_m$ varies from 9.20 to 9.17
using a 3-loop evaluation of $\alpha_s$.
Marciano \cite{Marc93} gives an updated value for $B_{\rm np}$
of $11.0 \pm 1.0$, but in view of the high momentum scales in
Eq.~(\ref{eq:MSnp}), suggests replacing $\alpha$ by
$\alpha(M_Z^2)$ in Eq.~(\ref{eq:boxgz}).
This value for $\Box_{\gamma Z}$ is the one adopted in
Ref.~\cite{Erler03}, and contributes almost half of the
error in the theoretical estimate $Q_W^p = 0.0713(8)$.

To progress in a systematic way beyond the approach of MS \cite{MS},
and to determine the dependence on energy $E$, we present a
new formulation of the box diagram contribution in which the dominant
part of the correction is expressed in terms of empirical moments of
structure functions.
%###
%
At forward angles one can compute $\Box_{\gamma Z}(E)$ from its
imaginary part using dispersion relations~\cite{GH}.
The imaginary part depends on the PV $ep \to eX$ cross section,
which can be expressed in terms of the product of leptonic and
hadronic tensors.
Following standard conventions \cite{PDG}, the hadronic tensor can be
written in terms of the interference electroweak structure functions as
\begin{equation}
M W^{\mu\nu}_{\gamma Z}
= - g^{\mu\nu} F_1^{\gamma Z}
  + {p^\mu p^\nu\over p\cdot q} F_2^{\gamma Z}
  - i \varepsilon^{\mu\nu\lambda\rho} {p_\lambda q_\rho \over 2 p\cdot q}
    F_3^{\gamma Z},
\end{equation}
where $p$ and $q$ are the four-momenta of the proton and exchanged
boson, respectively.
The $F_{1,2}^{\gamma Z}$ contributions to $\Box_{\gamma Z}$ involve
the vector hadron coupling of the $Z$, and were recently computed in
Refs.~\cite{GH,SBMT,RC,GHRM}.

Our focus here is on the $F_3^{\gamma Z}$ contribution involving
the axial-vector hadron coupling of the $Z$.
Following an analogous derivation in Ref.~\cite{SBMT}, we can write
\begin{eqnarray}
\Im{\rm m}\, \Box_{\gamma Z}^{\rm A}(E)
 &=& {1 \over (2 M E)^2}
     \int_{M^2}^s \!\!\! dW^2\int_0^{Q_{\rm max}^2}
     \!\!\! dQ^2					\quad\nonumber\\
 & & \hspace{-1.8cm}
     \times
     { v_e(Q^2)\, \alpha(Q^2)\, F_3^{\gamma Z} \over 1+Q^2/M_Z^2}
 \left({2 M E\over W^2-M^2+Q^2}-\frac{1}{2}\right),
\label{eq:imW}
 \end{eqnarray}
with $s=M^2+2 M E$ and $Q_{\rm max}^2=2 M E (1-W^2/s)$.
The real part is determined from the dispersion relation
\begin{equation}
\Re{\rm e}\, \Box_{\gamma Z}^{\rm A}(E)
= {2\over \pi} \int_0^\infty dE' {E'\over E'^2-E^2} \,
  \Im{\rm m}\,\Box_{\gamma Z}^{\rm A}(E'),
\label{eq:disp}
\end{equation}
which accounts for both the box and crossed-box terms.
Unlike the vector hadronic correction
$\Re{\rm e}\, \Box_{\gamma Z}^{\rm V}(E)$, which vanishes at $E=0$,
the axial-vector hadronic correction
$\Re{\rm e}\, \Box_{\gamma Z}^{\rm A}(E)$ remains finite, and is dominant
in atomic parity violation at very low electron energies \cite{APV}.

We incorporate one further improvement over earlier calculations by
allowing for the $Q^2$ dependence of $\alpha(Q^2)$
and $\sw(Q^2) = \kappa(Q^2)\, \hat{s}^2$ in Eq.~(\ref{eq:imW})
due to boson self-energy contributions.
Both quantities vary significantly over the range of $Q^2$
relevant to these integrals.
The photon vacuum polarization expression is well-known, and
expressions for the universal fermion and boson contributions
to $\kappa(Q^2)$ are given in Ref.~\cite{Czar00}.
Following Ref.~\cite{MS}, we use effective quark masses to reproduce the
hadronic contribution of $\Delta\alpha_{\rm had}^{(5)}(M_Z^2)=0.02786$
obtained from dispersion relations~\cite{PDG}, yielding
$\kappa(0)=1.030$.
This is sufficiently accurate for the purpose of calculating the box contributions.
In the numerical results that follow, the effect of using
$\alpha(Q^2)$ and $v_e(Q^2)$ reduces the
total contribution to Eq.~(\ref{eq:disp}) by 17\% relative
to using $\alpha$ and $v_e(M_Z^2)$.

The imaginary part of $\Box_{\gamma Z}^{\rm A}$ can be split
into three regions:
%###
(i) elastic (el) with $W^2=M^2$;
(ii) resonances (res) with $(M+m_\pi)^2\le W^2 \lesssim 4$~GeV$^2$;
and
(iii) deep inelastic (DIS), with $W^2>4$~GeV$^2$.
Contributions from region (i) can be written in terms of the elastic
form factors as
\begin{equation}
F_3^{\gamma Z (\rm el)}(Q^2)
= -Q^2 G_M^p(Q^2) G_A^Z(Q^2) \delta(W^2-M^2).
\end{equation}
For the proton magnetic form factor $G_M^p$ we use the recent
parametrization from Ref.~\cite{AMT} (the results are similar
if one uses a dipole with mass 0.84~GeV), and take the axial-vector
form factor to be
$G_A^Z(Q^2) = -1.267/(1+Q^2/M_A^2)^2$ with $M_A=1.0$~GeV.
A virtue of the dipole forms is that the integrals (\ref{eq:imW})
and (\ref{eq:disp}) can be performed analytically, which provides
a useful cross-check.

\begin{figure}[t]
\hspace*{-3mm}\includegraphics[width=7.5cm,clip=true,bb=-19 17 681 530]{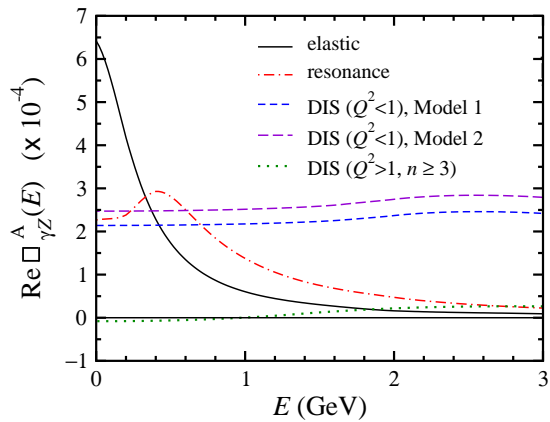}
\caption{Real part of $\Box_{\gamma Z}^{\rm A}(E)$ as a function of
	incident electron energy $E$. Shown are the elastic (solid)
	and resonance (dot-dashed) contributions. For the DIS part,
	the high-$Q^2$, $n \ge 3$ term (dotted) is negligibly small.
	The two $Q^2<1$~GeV$^2$ estimates (long and short dashes)
	show a very mild $E$ dependence. Not shown is the dominant
	high-$Q^2$, $n=1$ moment, which is $32.8\times 10^{-4}$,
	and is independent of $E$.}
\label{fig:ReBoxA}
\vspace*{-0.5cm}
\end{figure}

To simplify notation in what follows, we denote
$\Re{\rm e}\, \Box_{\gamma Z}^{\rm A}$ by $\Box_{\gamma Z}^{\rm A}$,
since that is the quantity of interest in Eq.~(\ref{eq:Qwp}).
The result for the elastic contribution
$\Box_{\gamma Z}^{\rm A(el)}(E)$ is shown in Fig.~\ref{fig:ReBoxA}.
It agrees exactly with the direct loop calculations of
$\Box_{\gamma Z}^{\rm A}$ in Refs.~\cite{Yang,TM},
in which the intermediate nucleon is off-shell.
It also agrees exactly at $E=0$ with the value $L_m=2.04$ if the
parameters are adjusted to correspond to those of MS~\cite{MS}.

For the resonance contributions $\Box_{\gamma Z}^{\rm A(res)}$ from
region (ii), we use the parametrizations of the transition form factors
from Lalakulich {\it et al.}~\cite{LPP}, but with modified isospin
factors appropriate to $\gamma Z$.  These form factors have been fitted
to the Jefferson Lab pion electroproduction data (vector part) and pion
production data in $\nu$ and $\bar\nu$ scattering at ANL, BNL and
Serpukhov (axial-vector part).  The parametrizations include the lowest
four spin-1/2 and 3/2 states in the first and second resonance regions,
up to $Q^2 = 3.5$~GeV$^2$.
At larger $Q^2$ the resonance contributions are suppressed by the
$Q^2$ dependence of the transition form factors, which is stronger
for the dominant $\Delta(1232)$ resonance than for the higher-mass 
resonances \cite{LPP}.
The resulting resonance contribution $\Box_{\gamma Z}^{\rm A(res)}(0)$
is smaller than the elastic term at $E=0$, but decreases less
rapidly with increasing energy.
Varying the $Q^2$ dependence of the poorly determined axial-vector form factors
has a negligible effect on these results.

To compute the DIS contributions from region (iii) it is convenient
to interchange the order of integration in (\ref{eq:imW}) and
(\ref{eq:disp}), in which case the integral over energy can be
performed analytically~\cite{RC}. A further change
of variable from $W^2$ to Bjorken $x=Q^2/(W^2-M^2+Q^2)$ gives
\begin{eqnarray}
\Box_{\gamma Z}^{\rm A(DIS)}(E)
&=& {2\over\pi}
    \int_{0}^\infty dQ^2
    {v_e(Q^2) \alpha(Q^2) \over Q^2(1+Q^2/M_Z^2)}	 \nonumber\\
& & \hspace*{-1.5cm}
    \times
    \int_0^{x_{\rm max}}\!\! dx\ F_3^{\gamma Z}(x,Q^2)\, f(r,t),
\label{eq:F3dis}					\\
& &\hspace*{-2.4cm}
f(r,t) = {1\over t^2} \left[ \log \left(1-t^2/r^2\right)
			   + 2 t \tanh^{-1} \left(t/r\right)
		      \right],				\nonumber
\end{eqnarray}
with $r \equiv 1 + \sqrt{1 + 4 M^2 x^2/Q^2}$, $t\equiv 4 M E x/Q^2$,
and $x_{\rm max} = Q^2/(W_{\rm min}^2-M^2+Q^2)$.
For $t=0$, we find $f(r,0)=(2 r-1)/r^2$.
In the free quark model limit with
$F_3^{\gamma Z}=(5/3)\,x\,\delta(1-x)$, Eq.~(\ref{eq:F3dis}) then
gives exactly the perturbative result of Eq.~(\ref{eq:MSp}) for $E=0$
(ignoring the $Q^2$ dependence of $\alpha$ and $v_e$).

To proceed, we divide the $Q^2$ integral of the full expression
(\ref{eq:F3dis}) into a low-$Q^2$ part, where the structure function
$F_3^{\gamma Z}$ is relatively unknown, and a high-$Q^2$ part
($Q^2 > Q_0^2$), where at leading order (LO) the structure functions
can be expressed in terms of valence quark distributions
$q_v = q-\bar{q}$~\cite{PDG},
\begin{equation}
F_3^{\gamma Z (\rm DIS)}(x,Q^2) = \sum_q 2\, e_q\, g_A^q\, q_v(x,Q^2).
\end{equation}
At high $Q^2$ and low $E$, the integrand in (\ref{eq:F3dis}) can be
expanded in powers of $x^2/Q^2$, yielding a series whose coefficients
are structure function moments of increasing rank,
\begin{subequations}
\label{eq:moments}
\begin{eqnarray}  
\label{eq:momA}
\Box_{\gamma Z}^{\rm A(DIS)}(E)
&=& {3\over 2\pi}
    \int_{Q_0^2}^\infty dQ^2
    {v_e(Q^2)\alpha(Q^2) \over Q^2(1+Q^2/M_Z^2)}  \nonumber\\
& & \hspace*{-3.1cm} \times
    \biggl[ M_3^{(1)}(Q^2)
	  + {2M^2 \over 9Q^4}(5E^2-3Q^2) M_3^{(3)}(Q^2) + \ldots
    \biggr].
\end{eqnarray}
For completeness, we also quote the result for the vector hadronic
correction,
\begin{eqnarray}
\label{eq:momV}
\Box_{\gamma Z}^{\rm V(DIS)}(E)
&=& {2 M E\over \pi}
    \int_{Q_0^2}^\infty dQ^2{\alpha(Q^2)\over Q^4(1+Q^2/M_Z^2)}   
						\nonumber\\
& & \hspace*{-2.5cm} \times
    \biggl[ M_2^{(2)}(Q^2)
          + \frac{2}{3} M_1^{(2)}(Q^2)
	  + {2M^2 \over 3Q^4} (E^2-Q^2) M_2^{(4)}(Q^2)	\nonumber\\
& & \hspace*{-2cm}
	  + {2M^2 \over 5Q^4} (4E^2 - 5Q^2) M_1^{(4)}(Q^2) + \ldots
    \biggr].
\end{eqnarray}
\end{subequations}
In Eqs.~(\ref{eq:moments}) the moments of the structure functions
are defined as
\begin{equation}
M_i^{(n)}(Q^2)
\equiv \int_0^1 dx\,x^{n-2} {\cal F}_i^{\gamma Z}(x,Q^2),\quad 
i=1,2,3,
\label{eq:momdef}
\end{equation}
where ${\cal F}_i^{\gamma Z}
       = \left\{ xF_1^{\gamma Z}, F_2^{\gamma Z}, xF_3^{\gamma Z}
	 \right\}$.
In approximating the upper limit $x_{\rm max}$ on the $x$-integrals
in Eqs.~(\ref{eq:moments}) by 1, the resulting error is less than
$10^{-4}$ for $Q^2 > 1$~GeV$^2$.
The large-$x$ contributions to $M_i^{(n)}(Q^2)$ become more important
for large $n$; however, the higher moments are suppressed by
increasing powers of $1/Q^2$.
In practice, the integrals in Eqs.~(\ref{eq:moments}) are dominated
by the lowest moments, with the $1/Q^2$ corrections being relatively
small in DIS kinematics.

Equations~(\ref{eq:moments}) are major new results which provide a
systematic framework within which to evaluate the radiative corrections.
For the axial-vector hadron part, the lowest moment, $M_3^{(1)}(Q^2)$,
is the $\gamma Z$ analog of the GLS sum rule \cite{GLS} for $\nu N$ DIS,
which at LO counts the number of valence quarks in the nucleon.
The corresponding quantity for $\gamma Z$ is
$\sum_q 2\, e_q\, g_A^q = 5/3$, so that at next-to-leading order
(NLO) in the $\overline{\rm MS}$ scheme
\begin{eqnarray}
M_3^{(1)}(Q^2)
&=& \!\!\frac{5}{3} \left( 1 - {\alpha_s(Q^2)\over \pi} \right), \\
M_3^{(3)}(Q^2)
&=& \!\!\frac{1}{3}
    \left(2 \langle x^2 \rangle_u + \langle x^2 \rangle_d \right)
    \left( 1 + {5 \alpha_s (Q^2)\over 12\pi} \right),\nonumber
\end{eqnarray}
where $\langle x^2 \rangle_q = \int_0^1 dx\, x^2\, q_v(x,Q^2)$.
Hence the lowest ($n=1$) moment contribution to Eq.~(\ref{eq:momA})
is identical to the MS result \cite{MS} in Eq.~(\ref{eq:MSnp}).
However, the parameter $Q_0^2$ in Eq.~(\ref{eq:momA}) has a
slightly different interpretation than the mass parameter $m^2$
of Eq.~(\ref{eq:MSnp}).
Here $Q_0$ corresponds to the momentum above which a partonic 
representation of the non-resonant structure functions is valid,
and above which the $Q^2$ evolution of parton distribution functions
(PDFs) {\it via} the $Q^2$ evolution equations is applicable.
We take $Q_0^2 = 1$~GeV$^2$, which coincides with the typical
lower limit of recent sets of PDFs \cite{MSTW,PDFs}.
The computation of the vector hadronic contribution to
$\Box_{\gamma Z}^{\rm (DIS)}$ proceeds in a similar manner,
and will be discussed elsewhere \cite{future}.

To evaluate the moments in Eq.~(\ref{eq:momA}) we use several NLO
parametrizations of PDFs determined from global fits
\cite{MSTW,PDFs}. The results are summarized in Fig.~\ref{fig:ReBoxA}.
Variations in the values of $\alpha_s(M_Z^2)$ among the datasets
considered had a negligible effect on the $n=1$ value of $0.0033$.
The $n=3$ moments for different datasets are virtually identical,
and give negligibly small contributions.

The $E$ dependent terms in Eq.~(\ref{eq:momA}) should also be small, 
since these depend on $n \geq 3$ moments.  However, the expansion in 
Eq.~(\ref{eq:momA}) is not strictly valid when $E > Q_0^2/2 M$.
To describe the $E$ dependence in this region we evaluate the difference
$\Box_{\gamma Z}^{\rm A(DIS)}(E) - \Box_{\gamma Z}^{\rm A(DIS)}(0)$
in Eq.~(\ref{eq:F3dis}) by replacing $f(r,t)$ by $f(r,t) - f(r,0)$.
The results are indeed small for $E$ in the few GeV region,
as the dotted line in Fig.~\ref{fig:ReBoxA} indicates.

For $Q^2 < Q_0^2$ a partonic description of the structure functions
is not valid.
In particular, since the integral over $Q^2$ in Eq.~(\ref{eq:F3dis})
extends down to $Q^2=0$, and the upper limit on the $x$-integral,
$x_{\rm max}$, is also limited by $Q^2$, one requires the behavior
of the structure functions at both low $x$ and low $Q^2$.
In the case of the vector $F_2^{\gamma Z}$ structure function,
conservation of the two vector currents requires
$F_2^{\gamma Z} \sim Q^2$ as $Q^2 \to 0$.
By contrast, $F_3^{\gamma Z}$ depends on both vector and axial-vector 
currents, and the nonconservation of the latter means that no similar 
constraint exists \cite{LPP}.

In the absence of data on $F_3^{\gamma Z}(x,Q^2)$ in the low-$x$,
low-$Q^2$ region, we consider models for the possible
$x$ and $Q^2$ dependence, obeying the following conditions:
 (1) $F_3^{\gamma Z}(x_\text{max},Q^2)$ should not diverge in
     the limit $Q^2\to 0$;
 (2) $F_3^{\gamma Z}(x,Q^2)$ should match the partonic structure
     function at $Q^2=Q_0^2$.
For the parametrization of Ref.~\cite{MSTW} we note that
$F_3^{\gamma Z}(x,Q_0^2) \sim x^{-0.7}$ as $x \to 0$.
With this in mind, we consider two models for $Q^2<Q_0^2$.

Model 1 sets
\begin{equation}
F_3^{\gamma Z}(x,Q^2)
= \left( { 1 + \Lambda^2/Q_0^2 \over 1 + \Lambda^2/Q^2} \right)
  F_3^{\gamma Z}(x,Q_0^2),
\end{equation}
which has the property that
$F_3^{\gamma Z}(x_\text{max},Q^2) \sim (Q^2)^{0.3}$
as $Q^2 \to 0$.
Here $\Lambda^2$ is a parameter that can be adjusted to examine
the model sensitivity of the integral in Eq.~(\ref{eq:F3dis}).
For $\Lambda^2$ in the range $(0.4-1.0)$~GeV$^2$,
we obtain a $\pm 10$\% variation in the values for
$\Box_{\gamma Z}^{\rm A}(E)$ shown in Fig.~\ref{fig:ReBoxA}.

Model 2 freezes $F_3^{\gamma Z}$ at the $Q^2=Q_0^2$ value
for all $W^2$, which is equivalent to setting
$F_3^{\gamma Z}(x,Q^2) = F_3^{\gamma Z}(x_0,Q_0^2)$,
with $x_0=x Q_0^2/\left((1-x)Q^2+x Q_0^2\right)$.
For this model, $F_3^{\gamma Z}$ is constant as $Q^2\to 0$,
and yields a 15\% larger contribution to
$\Box_{\gamma Z}^{\rm A}(E)$ than Model 1, as illustrated
in Fig.~\ref{fig:ReBoxA}.

\begin{figure}[t]
\hspace*{-1mm}\includegraphics[width=7.5cm]{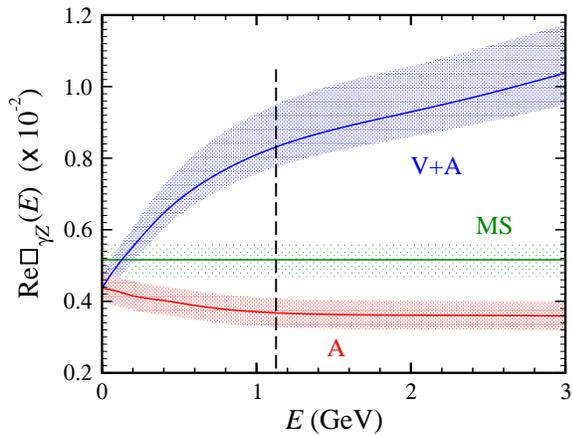}
\caption{Total (el+res+DIS) axial-vector hadron correction
	$\Box_{\gamma Z}^{\rm A}(E)$ (labeled ``A'')
	and the sum of axial and vector hadron \cite{SBMT}
	corrections (labeled ``V+A''), together with the 
	$E=0$ result of MS \cite{MS} (extended
	to finite $E$ for comparison).  The vertical dashed
	line indicates the energy at Q$_{\rm weak}$ kinematics.}
\label{fig:boxtotal}
\vspace*{-0.5cm}
\end{figure}

The total correction to $\Box_{\gamma Z}^{\rm A}$ is given by the sum
(el+res+DIS), and is shown in Fig.~\ref{fig:boxtotal} as a function
of $E$. As demonstrated, the $E$ dependence arises predominantly from
the elastic and resonance contributions.  We assign a very conservative
uncertainty estimate equal to twice the low-$Q^2$ DIS
value.
This allows for uncertainties in the resonance
and low-$Q^2$ DIS contributions, and in the effect of the running coupling
constants on the dominant $n=1$ contribution.
The total contribution to $\Box_{\gamma Z}^{\rm A}$ is
$0.0044(4)$ at $E=0$, and $0.0037(4)$ at $E=1.165$~GeV
(the Q$_{\rm weak}$ energy).
This should be compared to the value $0.0052(5)$ used in
Ref.~\cite{Erler03}, which is assumed to be energy independent.
Also shown in Fig.~\ref{fig:boxtotal} is the total $\Box_{\gamma Z}
= \Box_{\gamma Z}^{\rm V} + \Box_{\gamma Z}^{\rm A}$ using the result
for $\Box_{\gamma Z}^{\rm V}$ from Ref.~\cite{SBMT}, which has an
uncertainty that grows with $E$.

Our value shifts the theoretical estimate for $Q_W^p$ from $0.0713(8)$
to $0.0705(8)$, with a total energy dependent correction
$\Box_{\gamma Z}(E) - \Box_{\gamma Z}(0)$ of $0.0040^{+0.0011}_{-0.0004}$
at $E=1.165$~GeV.
A similar uncertainty would be obtained using the estimate of
$\Box_{\gamma Z}^{\rm V}$ from Ref.~\cite{RC}, while a larger
uncertainty on the vector hadron correction was quoted in
Ref.~\cite{GHRM}.
These uncertainties can be reduced with future PV structure function
measurements at low $Q^2$, such as those planned at Jefferson Lab.
The high precision determination of $Q_W^p$ would then allow more
robust extraction of signals for new physics beyond the Standard Model.

%%%%%%%%%%%%%%%%%%%%%%%%%%%%%%%%%%%%%%%%%%%%%%%%%%%%%%%%%%%%%%%%%%%%%%%%%
\begin{acknowledgments}
We thank C.~Carlson, O.~Lalakulich, E.~Paschos and A.~Sibirtsev for
helpful discussions.
This work is supported by NSERC (Canada), the DOE contract No.
DE-AC05-06OR23177, under which Jefferson Science Associates, LLC
operates Jefferson Lab, and the Australian Research Council through an
Australian Laureate Fellowship. PGB thanks Jefferson Lab and TRIUMF for
support during a sabbatical leave, where part of the work was completed.
\end{acknowledgments}

%%%%%%%%%%%%%%%%%%%%%%%%%%%%%%%%%%%%%%%%%%%%%%%%%%%%%%%%%%%%%%%%%%%%%%%%%

\end{document}